\documentclass[twocolumn]{svjour3}          
\smartqed  
\usepackage[T2A]{fontenc}
\usepackage[cp1251]{inputenc}
\usepackage[english]{babel}
\usepackage{amsmath,amssymb,amsfonts}

\usepackage{enumitem}
\usepackage{graphicx}
\usepackage[config, labelfont={sf,bf}, textfont=sf, caption=false]{subfig}
\usepackage{array}
\usepackage{multirow}
\usepackage{hhline}
\usepackage{color}
\usepackage{siunitx}

\usepackage[numbers,sort&compress]{natbib}

\usepackage[colorlinks,linkcolor=black,citecolor=blue,urlcolor=black]{hyperref}

\begin{document}\sloppy

\title{

  Emergence and evolution of unusual inhomogeneous limit cycles displacing hyperchaos in three quorum-sensing coupled identical ring oscillators
}

\author{ Nataliya Stankevich \and
         Evgeny Volkov
}


\institute{N. Stankevich \at
              Laboratory of Topological methods in Dynamics,\\
              HSE University, Nizhny Novgorod, Russia\\
              Kotel'nikov's Institute of Radio-Engineering and Electronics \\
              of RAS, Saratov Branch, Saratov, Russia\\
              \email{stankevichnv@mail.ru}
              \and
           E. Volkov \at
              Department of Theoretical Physics,\\
              Lebedev Physical Institute, Moscow, Russia\\
              \email{volkov@td.lpi.ru}
}

\date{January 30, 2022}

\maketitle

\begin{abstract}
We demonstrate that strongly asymmetric limit cycles can be observed in the system of three identical ring oscillators (3-gene networks known as Repressilators) globally coupled by signal molecule diffusion added to the model in a way like the known bacterial "quorum-sensing" mechanism. These cycles are stable over a wide interval of the coupling strengths where they expel the dominant hyperchaotic regime existing in three Repressilators in very large areas of parameters. 
The bifurcations of the inhomogeneous limit cycle, with a high-amplitude orbit for one oscillator and two low-amplitude identical orbits for the other two, are traced. Bifurcation analysis reveals an unusual cascade of bifurcations ended in the appearance of a new limit cycle with splitted (slightly nonidentical) low-amplitude orbits. Both cycles lose stability giving birth inhomogeneous chaos in the small parameter interval.
Hyperchaos dominates in the parameter plane around the "island" with inhomogenous limit cycles, and this accounts for very long hyper chaotic transients when the system is returning to stable asymmetric cycles after their perturbations. In turn, it is the cycles that contribute the asymmetric and often rather long pieces in hyperchaotic trajectories.  The presented cycles differ from the known asymmetric attractors: inhomogeneous limit cycles born from "oscillation death" and the cycles observing in "smallest chimeras".    \\
\end{abstract}

\section{Introduction}
\label{intro}
Models of coupled oscillators have been studied for many years in order to better understand complex phenomena in very different natural and artificial systems \cite{mosekilde2002chaotic, balanov2009simple, pikovsky2015dynamics, csaba2020coupled}. To determine their basic dynamic properties, the single oscillators exhibiting a limit-cycle attractor are usually considered. As generally accepted, the set of collective regimes is controlled by both the properties of single oscillators and of the design of their interactions. Hyperchaos is an interesting regime emerging, for example, in two chaotic R\"{o}ssler oscillators coupled by linear diffusion of one variable \cite{rasmussen1996bifurcations,postnov1999role,yanchuk2001chaos} and in two identical Chua circuits with bi-directional non-linear coupling \cite{cannas2002hyperchaotic} (as a few to mention among others). There are publications in which hyperchaos arises in coupled regular oscillators stimulated by external periodic oscillations that results in their chaotization \cite{kapitaniak1995chaos, stankevich2018chaos, garashchuk2019hyperchaos}.

Recently \cite{stankevich2021chaos}, we detected chaos-hyperchaos transition in the simplified model comprising three identical synthetic genetic ring oscillators. This oscillator, named Repressilator, became popular after its implementation in bacteria \cite{elowitz2000synthetic}. Stability of its dynamics was seriously improved after the revision made in \cite{niederholtmeyer2015rapid, potvin2016synchronous}. As a result Repressilator found interesting application \cite{riglar2019bacterial}.

The three genes in the ring produce mRNAs which are translated to proteins-repressors and they impose Hill function inhibition on each next gene in the ring (each ring member is inhibited by the preceding one and, in its turn, inhibits the next one). 
The simple ring structure of the Repressilator allows for the stability of 3-dimensional limit cycle and the absence of its period-doubling bifurcations over very large areas of control parameters.

Similar to other synthetic genetic networks, Repressilators implemented in different bacteria (or other appropriate containers impermeable to repressors), have to be specially connected via the external medium in order to form desired collective regimes. 
The mechanism of bacterial quorum sensing (QS) is a natural candidate for such a function, since it contains the necessary elements for the production of small specific molecules (autoinducers) that rapidly diffuse in the environment and can influence gene expression.
It has been explored in the model with frequency-detuned Repressilators \cite{garcia2004modeling} to demonstrate their complete in-phase synchronization as well as in the constructing in-phase regime \cite{mcmillen2002synchronizing} and chaos \cite{chen2011modeling} in quorum sensing coupled synthetic relaxation genetic oscillators. 

It is very important to underline the flexibility of the QS-coupling designs. Modern genetic engineering permits equipping a given gene with different promoters and different genes with the same promoter. 
For example, it is possible to put two genes under control of the identical promoter: the one gene codes autoinducer production and the second gene codes selected repressor.
The next essential possibility of modifying the QS-depending communications is more or less free choice of the gene in the ring to be targeted by the autoinducer. It is possible to add one more Repressilator's gene and control its expression by the promoter capable to accept the regulatory action of autoinducer.  
This flexibility of the coupling scheme opens new possibilities both for the generation of coherent behavior in heterogeneous populations \cite{mcmillen2002synchronizing, garcia2004modeling} and for the remarkable variability of collective regimes in homogeneous ones. One version has been developed in \cite{ullner2007multistability} and was intensively used to study very rich dynamics of two coupled identical 6-dimensional Repressilators \cite{ullner2008multistability, koseska2010cooperative}. 
As shown recently, in two QS-coupled 3-dimensional Repressilators there are wide regions of control parameters where stable strongly inhomogeneous limit cycle arises as a result of the unusual sequence of pitchfork and Neimark-Sacker bifurcations of the unstable in-phase limit cycle \cite{volkov2021effect}.     

The above mentioned results stimulate the investigation of collective regimes generated in three globally QS-coupled identical Repressilators. In general, three coupled identical oscillatory systems are the simplest ones capable of exhibiting frustration. Frustration plays an important role in the dynamics of three coupled oscillators because it may cause symmetry breaking that drives certain asymmetric synchronized modes. A group-theoretical model-independent analysis \cite{collins1994group} have presented several modes of oscillating patterns for three symmetrically coupled identical oscillators: (i) in-phase solution (InP); (ii) rotating wave (RW) \cite{ashwin1990three} (phase differences between the orbits are $2 \pi/3$): (iii) two in-phase orbits are in anti-phase with third one ("partial in-phase") and (iv) two orbits are in anti-phase providing for doubled frequency oscillation and small amplitude for the third oscillator (1anti1:1half-period).  

Several experimental reports demonstrate the entire or partial sets of these regimes. Yoshimoto et all \cite{yoshimoto1993coupling} observed InP, RW limit cycles in three tanks with the Belousov-Zhabotinsky reaction coupled by mass exchange and Vanag et al. \cite{vanag2000oscillatory} found the same regimes in a model of three identical Oregonators with global negative coupling. Limit cycles InP, RW and partial in-phase were observed as isolated or coexisting regimes in three electronic relaxation oscillators inhibitory coupled to a ring by ohmic resistors over a wide area of control parameters \cite{ruwisch1999collective}. Synchronization of three identical plastic bottle oscillators in a water bath coupled via two sets of different configurations of thin tubes also shows up as the emergence of the above mentioned regimes \cite{kohira2009synchronization}. 

Similar experiments with natural objects are not so easily controlled compared with the hand-made systems. However, different combinations of the discussed regimes are robustly observed. For example, three Plasmodium oscillators connected by protoplasmic streaming \cite{takamatsu2001spatiotemporal} showed RW, partial in-phase, and 1anti1half-period modes. When set along a straight line, three singing male Japanese tree frogs (which call females nearly periodically being isolated) demonstrate RW and 2in/1anti synchronization if they hear sounds including calls of other males \cite{aihara2011complex}.  
The collective behavior of three-coupled candle flame oscillators in a triangular arrangement showed three distinct types of synchronized modes \cite{okamoto2016synchronization} i.e. the in-phase mode, the partial in-phase mode, and the RW mode. In some experiments not all principally expected periodic regimes were observed. It is probably a result of the limited intervals of parameters scanning and/or of the deviations from symmetry in investigated systems. 

In most model networks cited above, interactions between the identical oscillators are realized using the same component taken from any oscillator and applied to other oscillators in the network with various weights which depend on the oscillator pairs that are coupled. As a rule, in the corresponding systems of ODEs such interactions are formalized as a classical diffusion term $D*(X(i+1)+X(i-1)-2X(i))$ in the equation for the chosen variable of the oscillator. Our version of QS-coupling is based on the diffusion of autoinducer which is not the principle component of the Repressilator limit cycle. Autoinducer production and target gene expression are associated with different Repressilator components ordered in time due to the ring structure. In addition, in our approach, coupling strength is controlled by dilution of signal molecules in the environment rather than by changing their diffusion coefficient. Therefore, we expect that the new attractors and unusual multistability might arise.

 As demonstrated in our previous publication \cite{stankevich2021chaos}, three QS-coupled Repressilators form a stable RW at very small coupling strengths. With an increase in the coupling strength, the RW cycle loses stability via Neimark-Sacker bifurcation, then torus destruction produces chaos which, in turn converts to hyperchaos after merging with the saddle RW cycle.
 Here, we aim to investigate the dynamics of three QS-coupled Repressilators over extended parameter intervals where hyperchaos has already formed. Tracing the evolution of hyperchaos we observe its replacement in a significant window of parameters by a strongly inhomogeneous limit cycle with winding numbers of 1:2:2 where one oscillator is running along a large-amplitude orbit out-of-phase with the two others, which move in-phase with small amplitudes ($\rm{LC\_1L2in2s}$). Detailed bifurcation analysis reveals an unusual set of inhomogeneous regimes inside the area of coupling strengths bounded by saddle-node bifurcations while the parameter intervals of their stability are controlled by the pitchfork, Neimark-Sacker, and period doubling bifurcations. Moreover, in the parameter areas where the discovered inhomogeneous limit cycles are unstable they manifest themselves as important elements of hyperchaotic skeleton.

The work is structured as follows. In Sec.~\ref{sec:1 Object}, we describe the model, the model parameters, the main steps to hyperchaos, classification of regimes according to the spectrum of Lyapunov exponents and the coarse graining map of regimes. In Sec.~\ref{sec:2 Res}, a detailed study of the system dynamics is given, with the focus on the evolution of the inhomogeneous limit cycles. In Section \ref{sec:Concl} the conclusions and the comparison of the revealed cycles with the other types of inhomogeneous cycles are presented.

\section{Object and model }
\label{sec:1 Object}

\begin{figure*}
\centering
\includegraphics[width=0.9\textwidth]{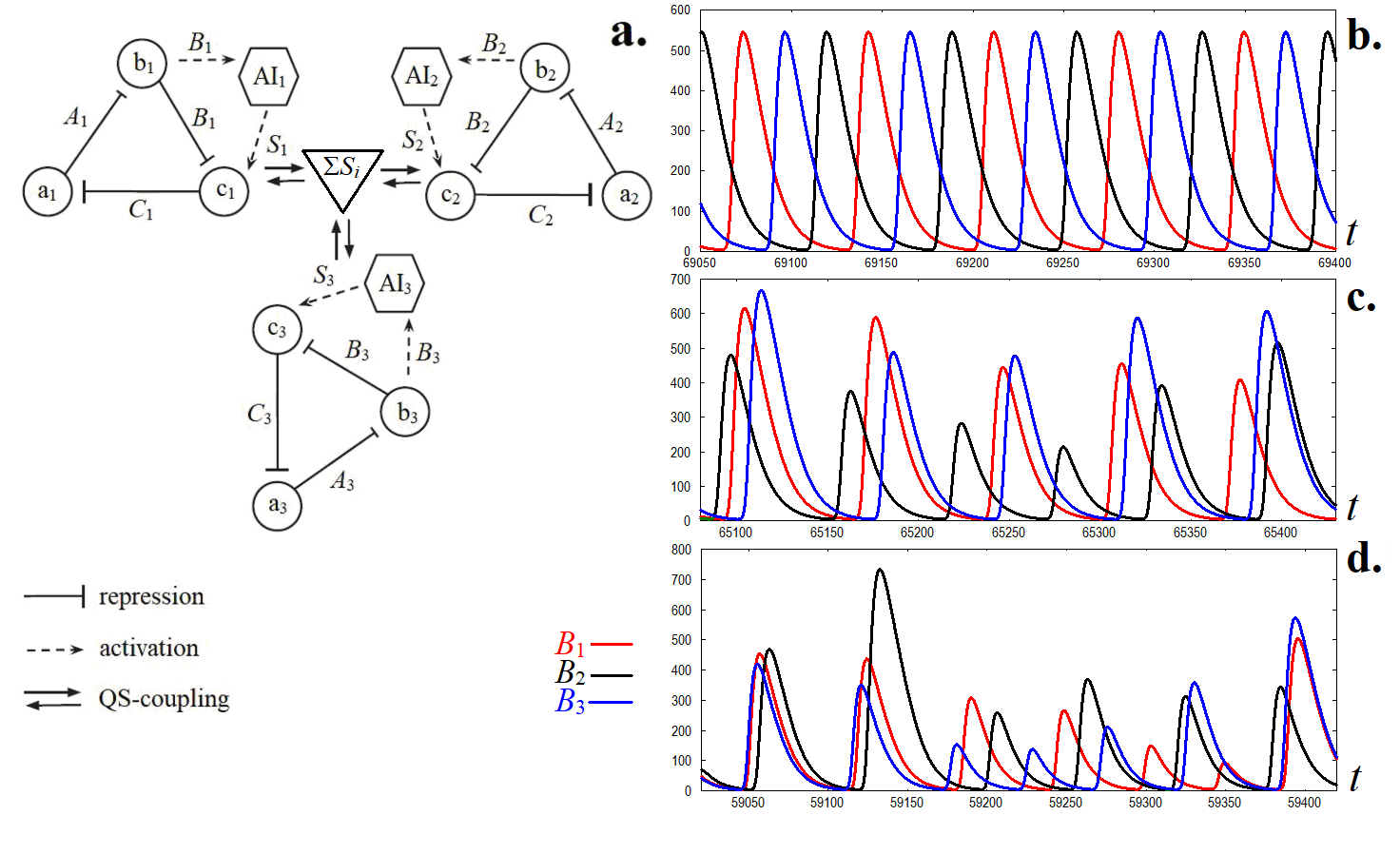}
\caption{\textbf{a.} Scheme of three Repressillators coupled via quorum-sensing mechanism; \textbf{b.}-\textbf{d.} Time series demonstrating main dynamical regimes: \textbf{b.} $Q=0.03$ - periodic regime (rotating wave); \textbf{c.} $Q=0.12$ - chaotic regime ($\Lambda_1 = 0.0011$, $\Lambda_2 = 0.0$, $\Lambda_3 = -0.0022$, $\Lambda_4 = -0.024$); \textbf{d.} $Q = 0.3$ - hyperchaotic regime ($\Lambda_1 = 0.0077$, $\Lambda_2 = 0.004$, $\Lambda_3 = 0.0$, $\Lambda_4 = -0.0259$). Other parameters as in (3) and $\alpha = 4000$}
\label{fig1:Scheme_3TS}
\end{figure*}

We investigate the dynamics of three Repressilators interacting via a bacterial quorum sensing (QS) mechanism used previously (see \cite{stankevich2021chaos} for details). Figure \ref{fig1:Scheme_3TS}a shows three Repressilators located in different cells and coupled via diffusion of special signal molecules (autoinducer, AI) in the external medium. 

The QS feedback is maintained by the AI produced (rate $k_{S1}$) proportionally to the protein $B$ concentration, while the autoinducer molecules diffuse quickly in the environment and activate (rate $\kappa$ in combination with Michaelis function) the production of mRNA for protein $C$, which, in turn, reduces the concentration of protein $A$, which results in the activation of protein $B$ production. In this way, protein $B$ plays a dual role of direct inhibition of protein $C$ synthesis and AI-dependent activation of protein $C$ synthesis, resulting in complex dynamics of the repressilator, even for a single repressilator. We reduce the model for the case of fast mRNA kinetics (($a$, $b$, $c$) are assumed in a steady state with their respective inhibitors ($C$, $A$, $B$), so that $da/dt=db/dt=dc/dt=0$). The resulting equations for dimensionless repressors ($A$, $B$, $C$) and autoinducer ($S$) concentrations are
\begin{equation}
\label{3Rep}
  \begin{array}{l l l l}
    \dot{A_i} = \beta_1 (-A_1 + \frac{\alpha}{1+C_i^n}),\\
    \dot{B_i} = \beta_2 (-B_1 + \frac{\alpha}{1+A_i^n}),\\
    \dot{C_i} = \beta_3 (-C_1 + \frac{\alpha}{1+B_i^n} + \frac{\kappa S_i}{1 + S_1}),\\
    \dot{S_i} = -k_{S0} S_i + k_{S1} B_i - \eta (S_i - S_ext),
  \end{array}
\end{equation}
where $i = 1, 2, 3$ are subscripts for the three repressilators, $\beta_j$ ($j = 1, 2, 3$) are the ratios of the protein decay rate to the mRNA decay rate, $\alpha$ stands for the maximum transcription rate in the absence of the inhibitor, and $n$ is the Hill cooperativity coefficient for inhibition. For the quorum-sensing pathway, $k_{S0}$ is the ratio of the $S$ decay rate to the mRNA decay rate, and as previously mentioned, $k_{S1}$ is the rate of production of $S$: $\kappa$ gives the strength of $S$-dependent activation of protein $C$ production. The diffusion coefficient $\eta$ depends on the membrane permeability to the $S$ molecule. The concentration of $S$ in the external medium $S_{ext}$ is determined (in a quasi-steady state approximation) by $S$ produced by all the repressilators ($S_1$, $S_2$, $S_3$) and a dilution factor $Q$,
\begin{equation}\label{Sext}
  S_{ext} = Q \frac{S_1 + S_2 + S_3}{3}.
\end{equation}
The model parameters are fixed identical for each repressilator and coincide with those proposed in \cite{stankevich2021chaos}: 
\begin{equation}\label{Param}
 \begin{array}{l l}
  \beta_1 = 0.5, \beta_{2,3} = 0.1, n=3, k_{S0} = 1, k_{S1} = 0.01,\\
  \eta = 2, \kappa = 15.
   \end{array}
\end{equation}
Here, we use the coupling strength $Q$ and the maximum transcription rate in the absence of inhibitor $\alpha$ as the control parameters. To recall the main steps leading to hyperchaos (for more details, see \cite{stankevich2021chaos}), Figures \ref{fig1:Scheme_3TS}b, \ref{fig1:Scheme_3TS}c, \ref{fig1:Scheme_3TS}d show time series for RW \cite{ashwin1990three}, chaos and hyperchaos that occur as the coupling force increases, in caption one can find Lyapunov exponents for chaotic time series.

Here we present a more in-depth study of collective regimes in the parameter areas where hyperchaos dominates. As a first step we calculate the spectrum of Lyapunov exponents over the coarse graining of the ($Q$ - $\alpha$) parameter plane. Model (\ref{3Rep}) has 12-dimensional phase space and is characterized by 12 Lyapunov exponents, but only the largest forth exponents are important and used for regime classification presented in Table \ref{tab:LEs}. 

\begin{table}
\caption{Dynamic mode descriptors used in the chart of Lyapunov exponents}
\label{tab:LEs}       
\begin{tabular}{cccc}
\hline\noalign{\smallskip}
Dynamic regime & Label & Color & Spectrum of LEs \\
\noalign{\smallskip}\hline\noalign{\smallskip}
periodic oscillations & P & red & $\Lambda_1=0$,\\
 &  &  & $0>\Lambda_{2}>\Lambda_{3}>\Lambda_{4}$ \\
\hline
two-frequency & T2 & yellow & $\Lambda_{1,2} = 0$, \\
quasiperiodic oscillations &  & & $0>\Lambda_{3}>\Lambda_{4}$ \\
\hline
three-frequency & T3 & blue & $\Lambda_{1, 2, 3} = 0$ \\
quasiperiodic oscillations & & & $0>\Lambda_{4}$ \\
\hline
chaotic oscillations & C & grey & $\Lambda_1 > 0$, $\Lambda_2 = 0$, \\
 & & & $0>\Lambda_{3}>\Lambda_{4}$\\
\hline
hyperchaotic oscillations & HC & white & $\Lambda_1 > \Lambda_2 >0$,\\
 &  &  & $\Lambda_3 = 0$, $0>\Lambda_{4}$ \\
\noalign{\smallskip}\hline
\end{tabular}
\end{table}

\begin{figure}[h]
\begin{center}
\includegraphics[scale=0.85]{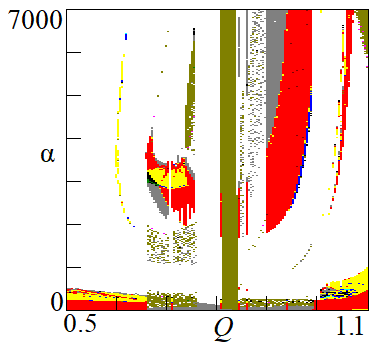}
\end{center}
\caption{The map of regimes classified according definitions described in Table1. Olive color points presented on the chart means that for these parameters the spectrum of Lyapunov exponents do not coincide with classification presented in Table \ref{tab:LEs}. The reasons of these specific points will be discussed below.
}
\label{fig2:chartLEs}
\end{figure}

The chart shown in Fig.~\ref{fig2:chartLEs} was constructed by scanning the parameter plane using adiabatic initial conditions and increasing parameter $\alpha$ in a stepwise manner: we take fixed random initial conditions for each new step of $Q$ value at $\alpha = 0$, and then for each new value of $\alpha$, we take initial conditions from the last step of the previous $\alpha$. Such calculations do not give a complete analysis of the dynamics but show the general layout of the mode map for a further detailed study. 
The sequence of the stable attractors as a function of $Q$ for small $\alpha < 500$ is the same as was observed in our previous paper \cite{stankevich2021chaos}: RW, quasiperiodic oscillation after Neimark-Sacker bifurcation of RW accompanied by resonant cycles formation, chaotic dynamics followed by hyperchaos formation.

After expanding the parameter plane, our parameter scanning showed that there are significant intervals of $\alpha$ and $Q$ values that form an "island" with limit cycles and collective modes  arising after their bifurcations. Their detailed study and evolution over the entire plane of ($\alpha$ - $Q$) parameters is the main goal of the current work.
The numerical results of the considered system are obtained through the Runge-Kutta fourth-order scheme, and the bifurcation analysis is carried out using the XPPAUT software \cite{ErmentroutXPPAUTO-2002}.

\section{Results}
\label{sec:2 Res}
\begin{figure}[h]
\begin{center}
\includegraphics[scale=0.5]{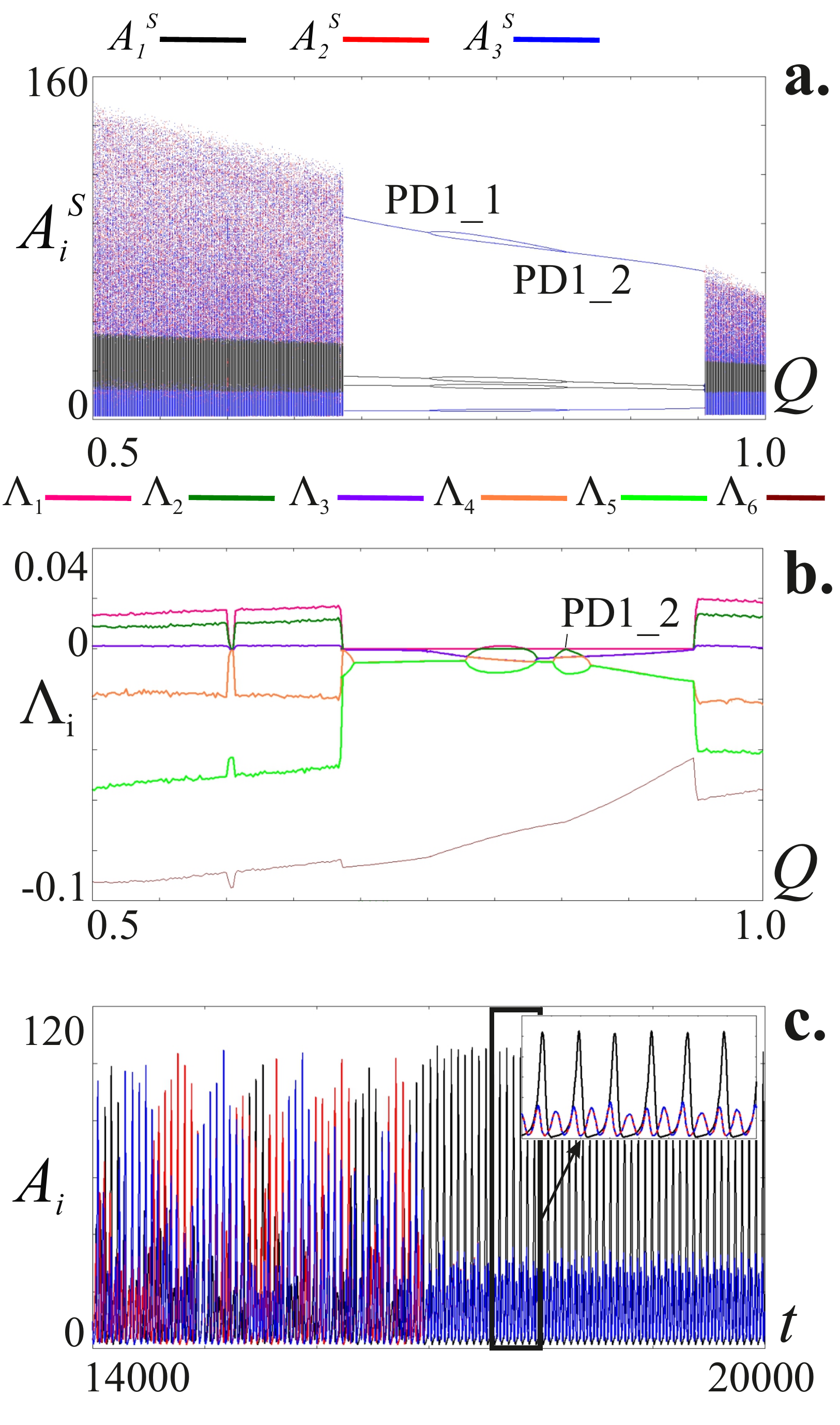}
\end{center}
\caption{Three Repressilators demonstrate the transition hyperchaos to inhomogeneous limit cycle (LC1L2in2s) vs the coupling strength $Q$ for $\alpha = 2777$: \textbf{a.} bifurcation tree obtained by direct integration with Poincar\'{e} section $C_1 = 7.0$; \textbf{b.} The six largest Lyapunov exponents; \textbf{c.} long time series, including the moment of transition and short inset with orbit for $Q = 0.686$.   
}
\label{fig3:SR1}
\end{figure}

We choose $\alpha = 2777$ for detailed analysis of hyperchaos evolution. In Fig.~\ref{fig3:SR1} bifurcation tree (\ref{fig3:SR1}a), plots of Lyapunov exponents (\ref{fig3:SR1}b) as a function of coupling strength $Q$ are presented. The bifurcation tree was constructed using the points (marked by $A_1^S$, $A_2^S$, $A_3^S$) in the cross-section of phase space by hypersurface $C_1 = 7.0$. At the start, the initial conditions were fixed as ($A_1^0 = 34.62$, $B_1^0 = 6.6$, $C_1^0 = 5.0$, $S_1^0 = 0.12$, $A_2^0 = 4.7$, $B_2^0 = 25.3$, $C_2^0 = 7.6$, $S_2^0 = 0.18$, $A_3^0 = 34.2$, $B_3^0 = 9.7$, $C_3^0 = 4.2$, $S_3^0 = 0.13$), and then we use stepwise continuation of parameter $Q$ changing the initial conditions after each step: the values of last phase variables from the previous step are used as initial conditions for the next step. The bifurcation tree and plots of Lyapunov exponents were calculated with different direction of scanning the $Q$-intervals: $Q$[0.5-0.75] and $Q$[1.0-0.75]. Fig.~\ref{fig3:SR1}c shows the time series with transition to limit cycle and transient process.

The bifurcation trees, Lyapunov exponents and the times series show the transition hyperchaos into inhomogeneous limit cycle marked as LC1L2in2s because two oscillators with small amplitudes (labelled as s) are running in-phase being out of phase with the large-amplitude (labelled as L) third oscillator as seen in the zoomed part of the time series (Fig.~\ref{fig3:SR1} inset).

\begin{figure*}
\includegraphics[width=0.7\textwidth]{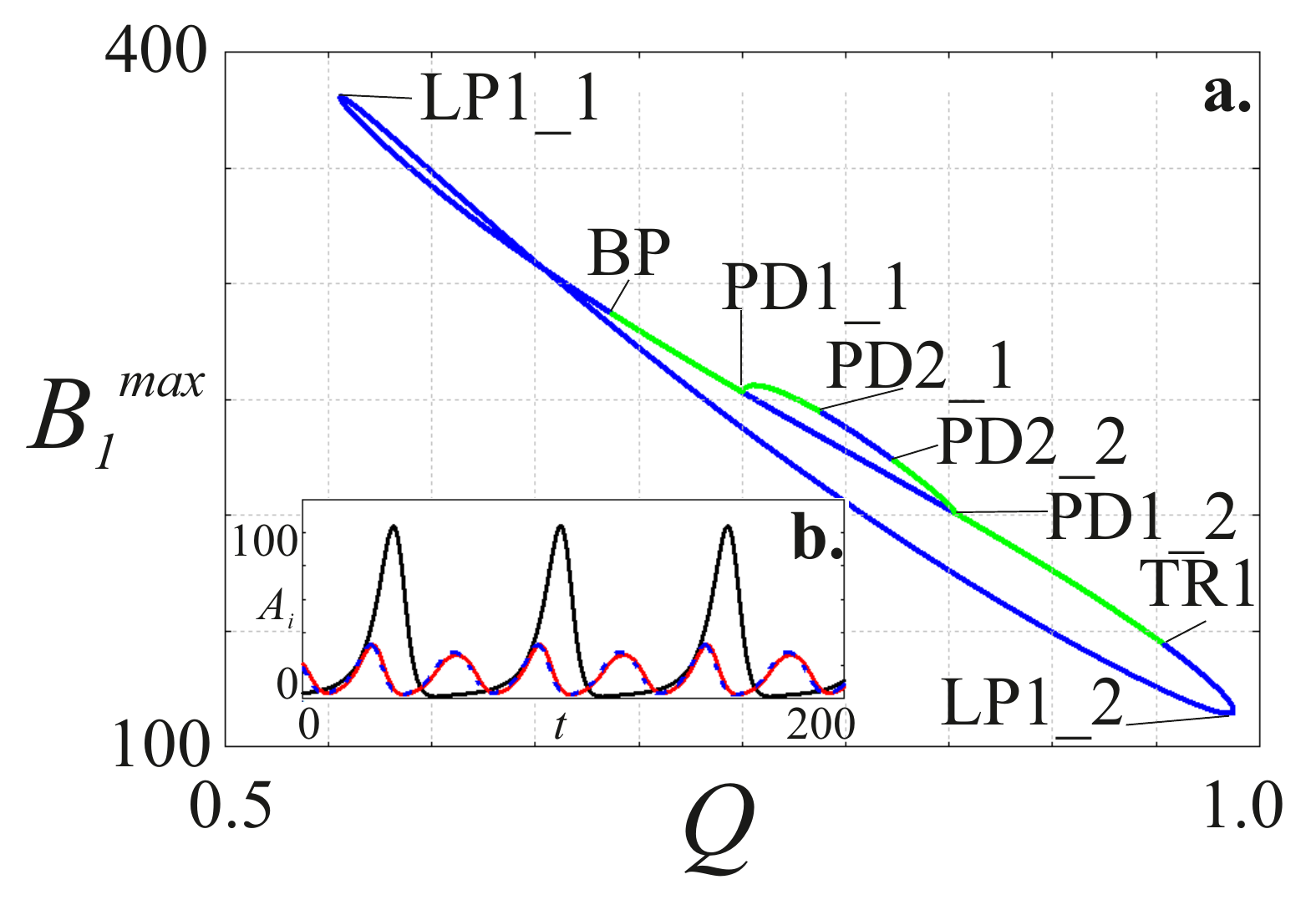}
\caption{\textbf{a.} AUTO continuation of LC1L2in2s: stable (unstable) part is green (blue) vs $Q$. Bifurcations: $Q = 0.55$ - saddle-node (LP1\_1); $Q = 0.68597$ - pitchfork (BP), $Q = 0.75$ - period doubling (PD1\_1); $Q = 0.789$ - PD2\_1; $Q = 0.823$ - PD2\_2; $Q = 0.8526$ - PD1\_2; $Q = 0.9542$ - Neimark-Sacker (TR1); $Q = 0.9876$ - LP1\_2. The regimes after the period doubling bifurcation PD2\_1, PD2\_2 will be considered below. \textbf{b.} Time series of the unstable LC1L2out2s just after the switching of LC1L2in2s continuation in the point of the pitchfork bifurcation at $Q = 0.68597$ 
}
\label{fig4:SR2_XPP2777}
\end{figure*}

\begin{figure*}
\includegraphics[width=0.7\textwidth]{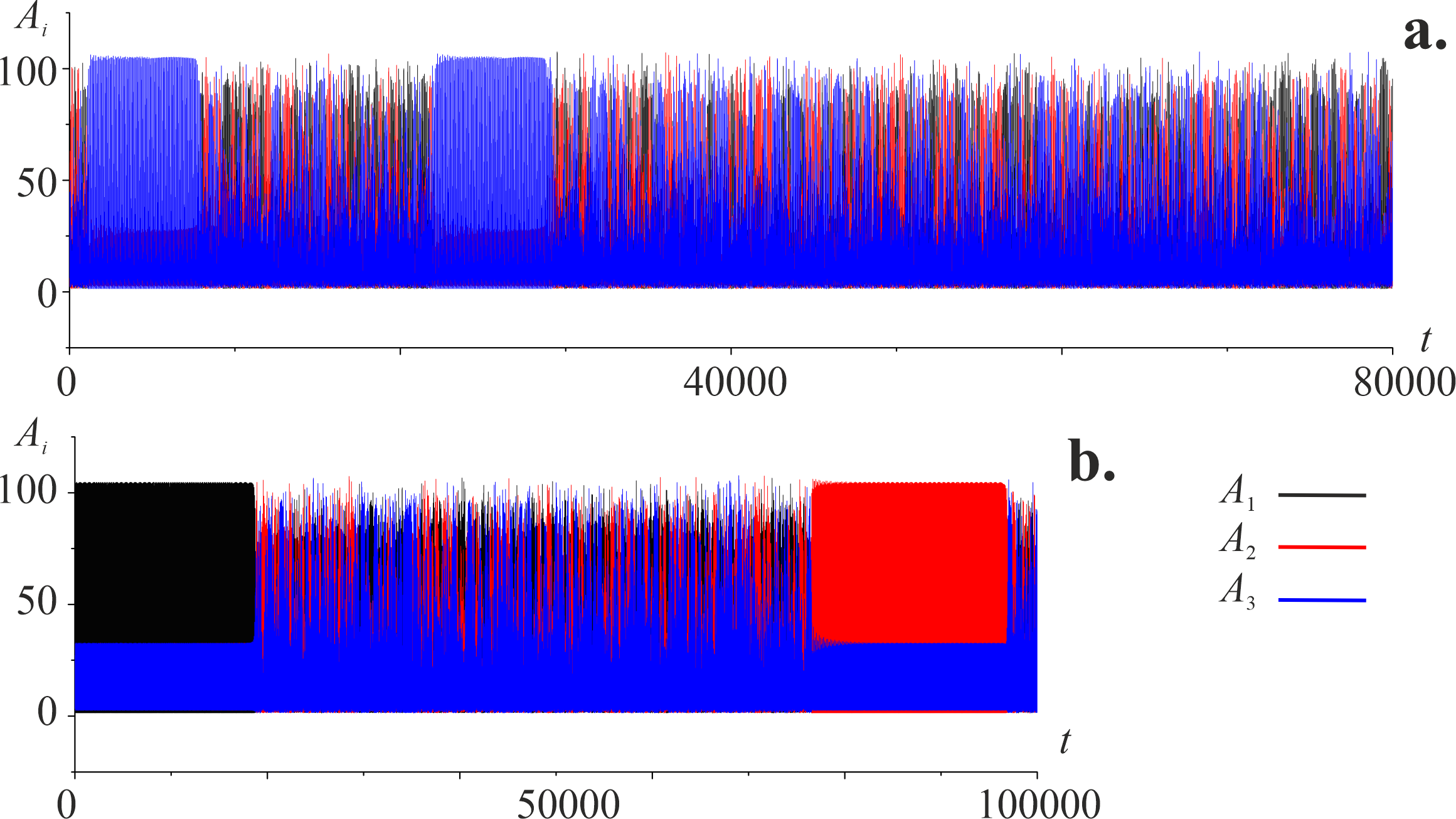}
\caption{Time series ($A_1$, $A_2$, $A_3$) show the intermittency between hyperchaos and LC1L2in2s: \textbf{a.} $Q = 0.68265$, \textbf{b.} $Q = 0.68545$
}
\label{fig5:SR3_TSs}
\end{figure*}

The detected cycle is stable with respect to external perturbations that destroy the in-phase regime of two elements. Bifurcation analysis by XPPAUT package \cite{ErmentroutXPPAUTO-2002} showed its nontrivial dynamics in a broad interval of the coupling strengths, see Fig.~\ref{fig4:SR2_XPP2777}.

\begin{figure*}
\includegraphics[width=0.7\textwidth]{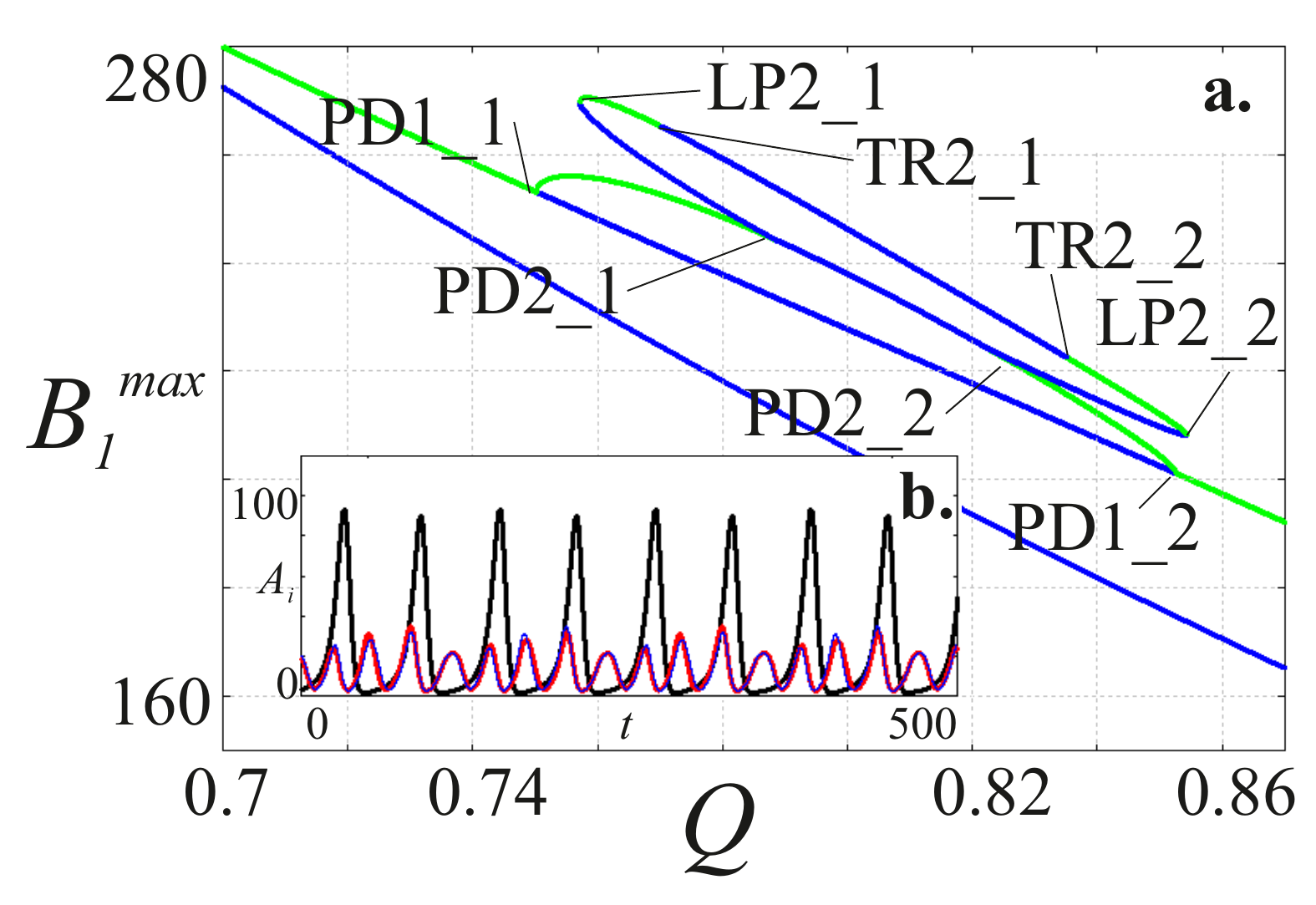}
\caption{The bifurcation continuations of inhomogeneous limit cycles starting from LC1L2in2s: the lines corresponding to stable (unstable) regimes are green (blue), respectively. LP2\_1 - $Q = 0.757$; PD1\_1 - $Q = 0.75$; TR1\_1 - $Q = 0.77$; PD2\_1 - $Q = 0.789$; TR1\_2 - $Q = 0.835$; PD2\_2 - $Q = 0.823$; LP2\_2 - $Q = 0.854$; PD1\_2 - $Q = 0.8526$. See text for the explanation. Inset \textbf{b.}: Unstable time series of LC4L8out8s extracted in the course of AUTO-continuation at $Q = 0.7675$
}
\label{fig7:SR5_XPP2777_Zoom}
\end{figure*}

Two saddle-node bifurcations LP1\_1 and LP1\_2 are the external boundaries of the interval within which the LC1L2in2s exists. But this interval doesn't coincide with interval where we can see manifestations of the limit cycle with direct integration (Fig.~\ref{fig3:SR1}a, \ref{fig3:SR1}b), because it isn't always stable between LP\_1 and LP\_2. This interval includes regions where other limit cycles are also stable and can be analyzed by direct integration. The internal boudaries of the parameter intervals where LC1L2in2s is stable are revealed by XPPAUT bifurcation continuation as a consequence of pitchfork (BP, $Q = 0.686$), period doubling and Neimark-Sacker (TR1, $Q = 0.9542$) bifurcations as presented in Fig.~\ref{fig4:SR2_XPP2777}.  The switching of the stable LC1L2in2s continuation in the BP-point results in the appearance of new unstable inhomogeneous limit cycle. The line of its bifurcation continuation is mainly overlapped with that for stable LC1L2in2s and therefore not shown in Fig.~\ref{fig4:SR2_XPP2777}. Stable LC1L2in2s is already strongly asymmetric (see Fig.~\ref{fig3:SR1}c inset) and its pitchfork bifurcation is manifested as the delicate splitting of two identical small-amplitude orbits observed in the zoomed time series (Fig.~\ref{fig4:SR2_XPP2777}b inset) extracted by XPPAUT option which permits such visualization. 

Although the both branches of LC1L2in2s are unstable between bifurcations LP1\_1 and BP, their manifestations in the form of intermittency with hyperchaos is often clearly observed in the narrow $Q$-interval very close to pitchfork bifurcation at $Q = 0.68597$, see typical examples in Fig.~\ref{fig5:SR3_TSs}. 

For coupling strengths greater than the value corresponding to BP bifurcation, the LC1L2in2s is the only stable attractor over the $Q$-interval up to its period doubling (PD1\_1) but hyperchaos can manifest itself as transients if the inhomogeneous cycle is perturbed. The durations of transients are often very long depending on the choice of the initial points. 

A further increase in coupling strength reveals the interesting cascade of LC1L2in2s bifurcations: period doubling bifurcation at $Q = 0.75$ produces a new stable branch of the cycle LC2L4in4s continuation, which, in turn, met the next period doubling bifurcation at $Q = 0.789$ (see Fig.~\ref{fig7:SR5_XPP2777_Zoom}).

\begin{figure}[h]
\begin{center}
\includegraphics[scale=0.55]{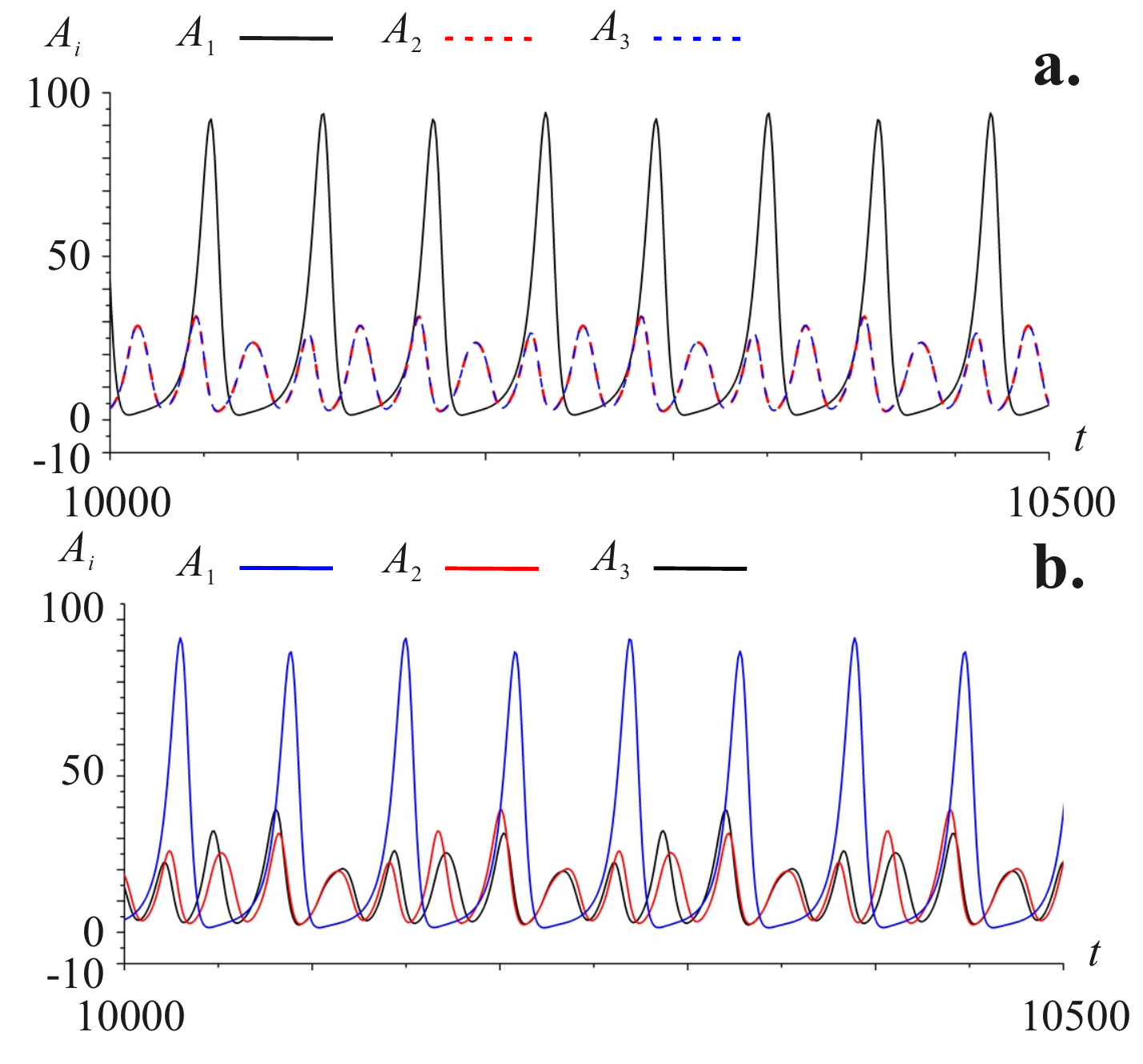}
\end{center}
\caption{Time series of stable inhomogeneous limit cycles at $Q = 0.762$: \textbf{a.} LC2L4in4s, \textbf{b.} LC4L8out8s
}
\label{fig8:SR6_TS}
\end{figure}

However, in contrast to the expected stable homogeneous LC4L8in8s, this bifurcation results in the appearance of an unstable inhomogeneous limit cycle with winding numbers 4L:8out8s, in which small-amplitude orbits are not in the in-phase regime any longer as seen in Fig.~\ref{fig7:SR5_XPP2777_Zoom}b. 
We assume that the presence of two oscillators with in-phase identical orbits in 12-dimensional phase space complicates the analysis of bifurcations of LC2L4in4s in this bifurcation point by the AUTO package. We observe, that the bifurcation identification by XPPAUT depends on the direction of the continuation along the $Q$-value and on the limit cycle selected for the continuation. In particular, PD2\_1 bifurcation at $Q = 0.789$ is combined with pitchfork bifurcation which splits two identical trajectories as seen in Fig.~\ref{fig7:SR5_XPP2777_Zoom}b. Figuratively speaking, this bifurcation occurs in 8-dimension subspace and qualitatively changes the type of the whole attractor.
Depending on these conditions, XPPAUT gives three versions for bifurcations PD2\_1 at $Q = 0.789$ and PD2\_2 at $Q = 0.823$ (Fig.~\ref{fig7:SR5_XPP2777_Zoom}), namely: period doubling, pitchfork and Neimark-Sacker bifurcations. The realizations all of them may be observed as the unstable time series extracted from the XPPAUT-continuation after these complex bifurcations. The new unstable LC4L8out8s is a result of three steps: (i) period doubling provides winding numbers 488; (ii) the coincidence of the small-amplitude trajectories of two Repressilators is broken via pitchfork formation; (iii) stability, expected after period doubling of the previous stable cycle, disappears due to the torus formation. 
The backward turn of this LC4L8out8s $Q$-continuation branch shows its instability up to the saddle-node bifurcation LP2\_1 ($Q = 0.757$) where it becomes stable leading to the coexistence of two stable inhomogeneous limit cycles, 2L4in4s, 4L8out8s, located on the ends of two branches of AUTO continuation over different intervals of the coupling strength (Fig.~\ref{fig7:SR5_XPP2777_Zoom}). The first LC2L4in4s is born in PD1\_1 bifurcation at $Q = 0.75$; the second LC4L8out8s, in complex PD2\_1-BP bifurcation, being stabilized in saddle-node LP2\_1 bifurcation at $Q = 0.757$. 
Both orbits are presented in Fig.~\ref{fig8:SR6_TS}.

\begin{figure}[h]
\begin{center}
\includegraphics[scale=0.7]{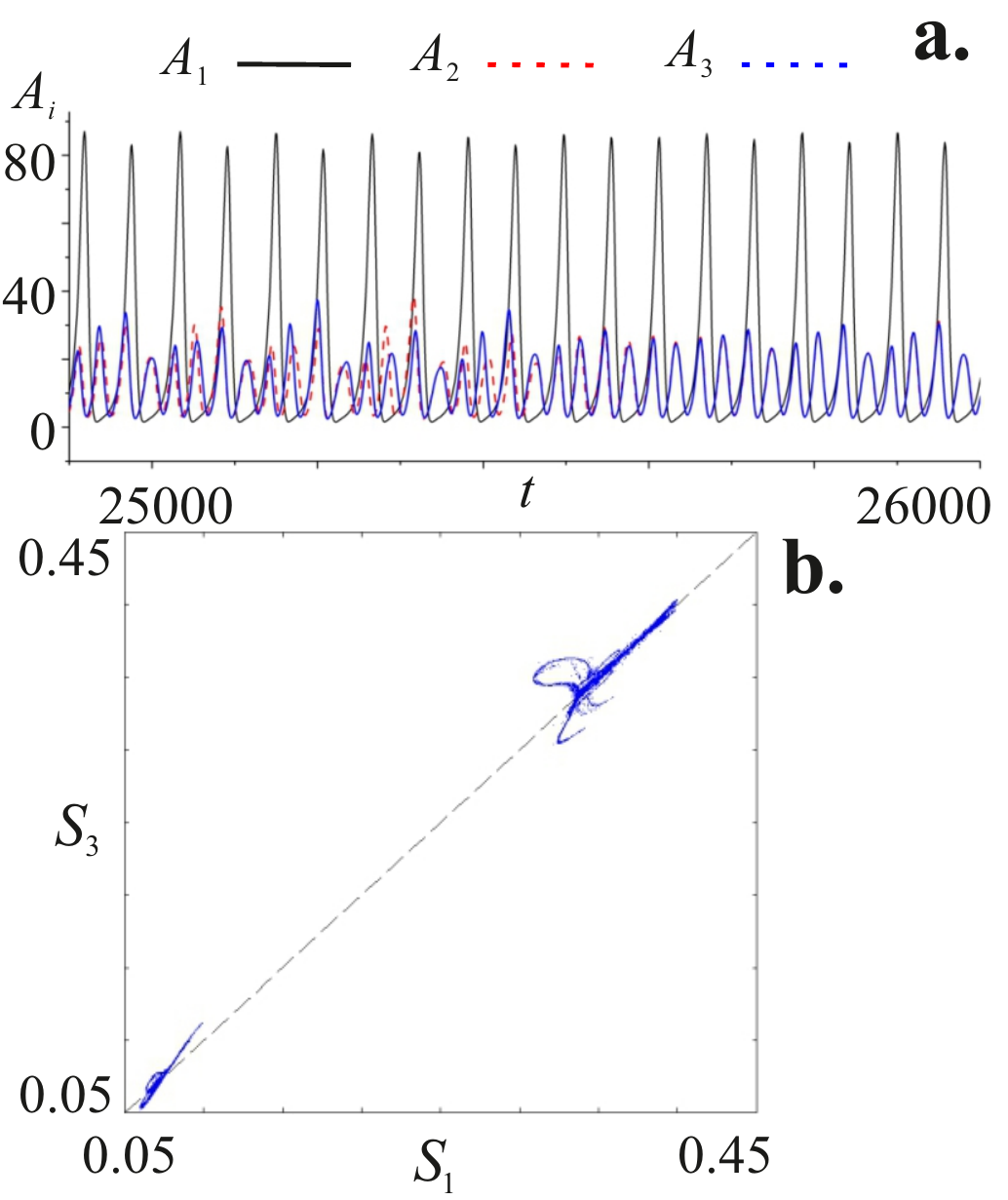}
\end{center}
\caption{\textbf{a.} Time series of attractor for $\alpha = 2777$, $Q = 0.808$; \textbf{b.} - Poincar\'{e} section
}
\label{fig9:SR10_TS_PS}
\end{figure}

There is a small region of coupling strength around $Q = 0.8$ where all inhomogeneous limit cycles are unstable and get mixed up in the phase space. By way of example, this regime is depicted in Fig.~\ref{fig9:SR10_TS_PS}a, \ref{fig9:SR10_TS_PS}b. Figure ~\ref{fig9:SR10_TS_PS}a shows randomly chosen time series $A_1$, $A_2$, $A_3$ with the small amplitude variables differing in one part and coinciding in another part. As can be seen from a Poincar\'{e} section of the long orbit by hypersurface $C_1 = 5.0$ given in Fig.~\ref{fig9:SR10_TS_PS}b, some points are located along bisectors $S_1 = S_3$ demonstrating existence of time intervals with in-phase oscillations. Other points deviate from bisectors revealing the time series intervals where small-amplitude oscillations are different. This map confirms that the structure of short trajectories in Fig.~\ref{fig9:SR10_TS_PS}a is typical of this attractor. Its first Lyapunov exponent is equal to 0.002 indicating that the inhomogeneous chaotic regime is too weak to stimulate the reappearance of hyperchaos.


As the coupling strength increases, the stability of inhomogeneous limit cycles is restored by means of the same bifurcations in which they were born: cycles 2LC4in4s, 4LC8out8s and LC1L2in2s become stable at $Q = 0.823$, $Q = 0.835$ and at $Q = 0.8526$, respectively (see right part of Fig.~\ref{fig7:SR5_XPP2777_Zoom}). The LC1L2in2s loses stability via Neimark-Sacker bifurcation at $Q = 0.9542$ (Fig.~\ref{fig4:SR2_XPP2777}) restoring the full dominance of hyperchaos in the system. The final disappearance of unstable LC1L2in2s occurs via saddle-node bifurcation at $Q = 0.9876$.

\begin{figure}
\begin{center}
\includegraphics[scale=0.45]{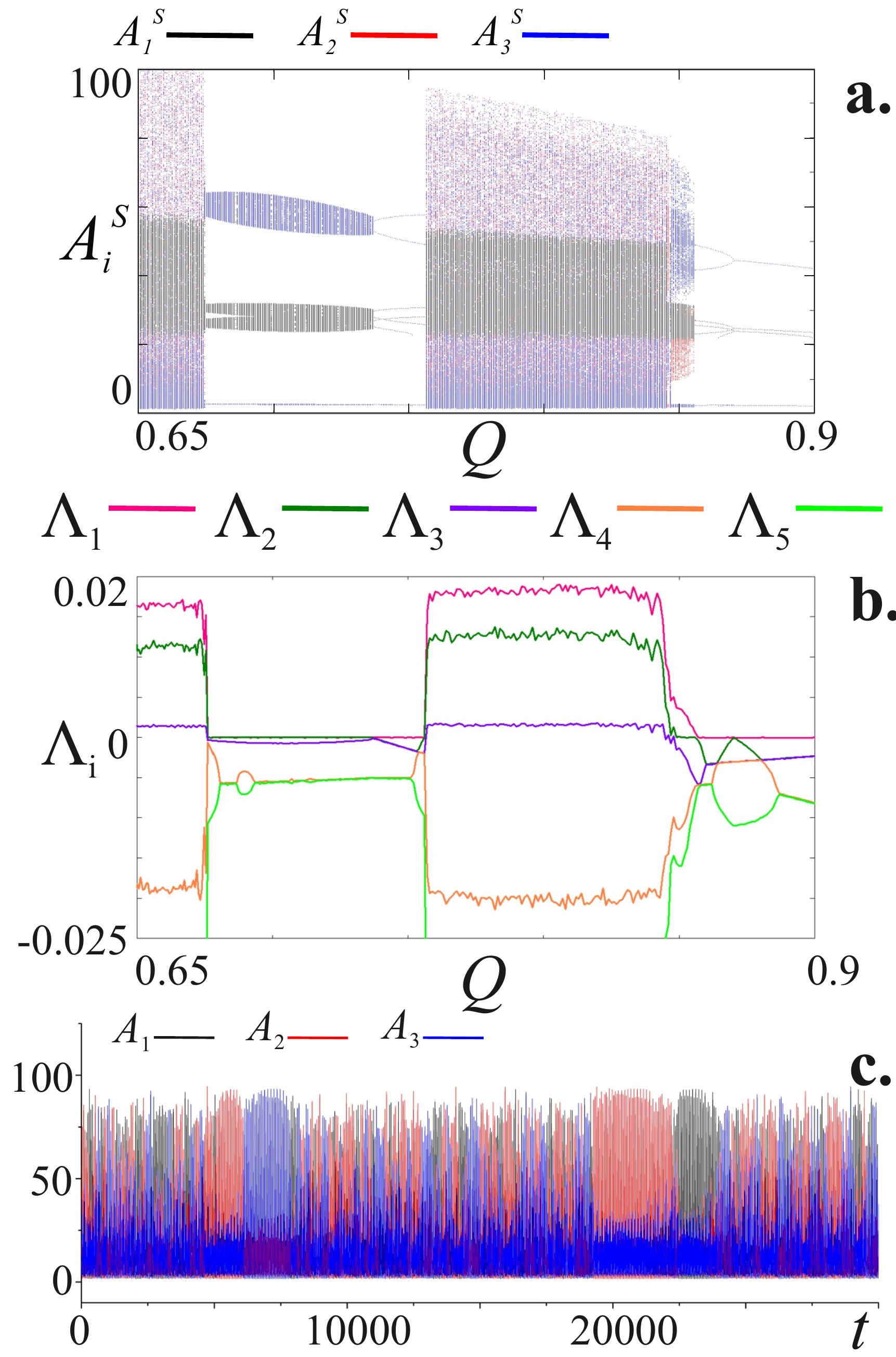}
\end{center}
\caption{Three Repressilators demonstrate the transition hyperchaos to inhomogeneous torus vs the coupling strength $Q$ for $\alpha = 3000$: \textbf{a.} bifurcation tree obtained by direct integration with Poincar\'{e} section $C_1 = 5.0$; \textbf{b.} The five largest Lyapunov exponents; \textbf{c.} long time series, including the moment of transition and short inset with orbit for $Q = 0.787$.   
}
\label{fig10:SR8_BT_LEs_TS}
\end{figure}

The interesting and complex dynamic demonstrated for $\alpha = 2777$ over the interval of coupling strengths between two saddle-node bifurcations (LP1\_1, LP1\_2) of LC1L2in2s undergo transformation when $\alpha$ is varied. The map of regimes in Fig.~\ref{fig2:chartLEs} shows the torus formation for $\alpha > 2777$ and $Q$ around 0.7.  The hyperchaos transition to inhomogeneous torus rather than LC1L2in2s at $Q = 0.675$ was traced by calculating the bifurcation tree and Lyapunov exponents for $\alpha = 3000$ (see Fig.~\ref{fig10:SR8_BT_LEs_TS}). 

The appearance of the inhomogenous quasiperiodic regime is a qualitative change in the hyperchaos evolution after $\alpha$ is increased slightly. As $Q$ increases the torus is replaced by LC1L2in2s and further dynamics of limit cycles is very similar to that pictured in Figs.~\ref{fig4:SR2_XPP2777}, \ref{fig7:SR5_XPP2777_Zoom}. When $Q$ is varied in the interval [0.77-0.84], the cycles lose stability and hyperchaos is back; however, it alternates with wide stretches of unstable time series of inhomogeneous limit cycles (see Fig9c).

\begin{figure}[h]
\begin{center}
\includegraphics[scale=0.2]{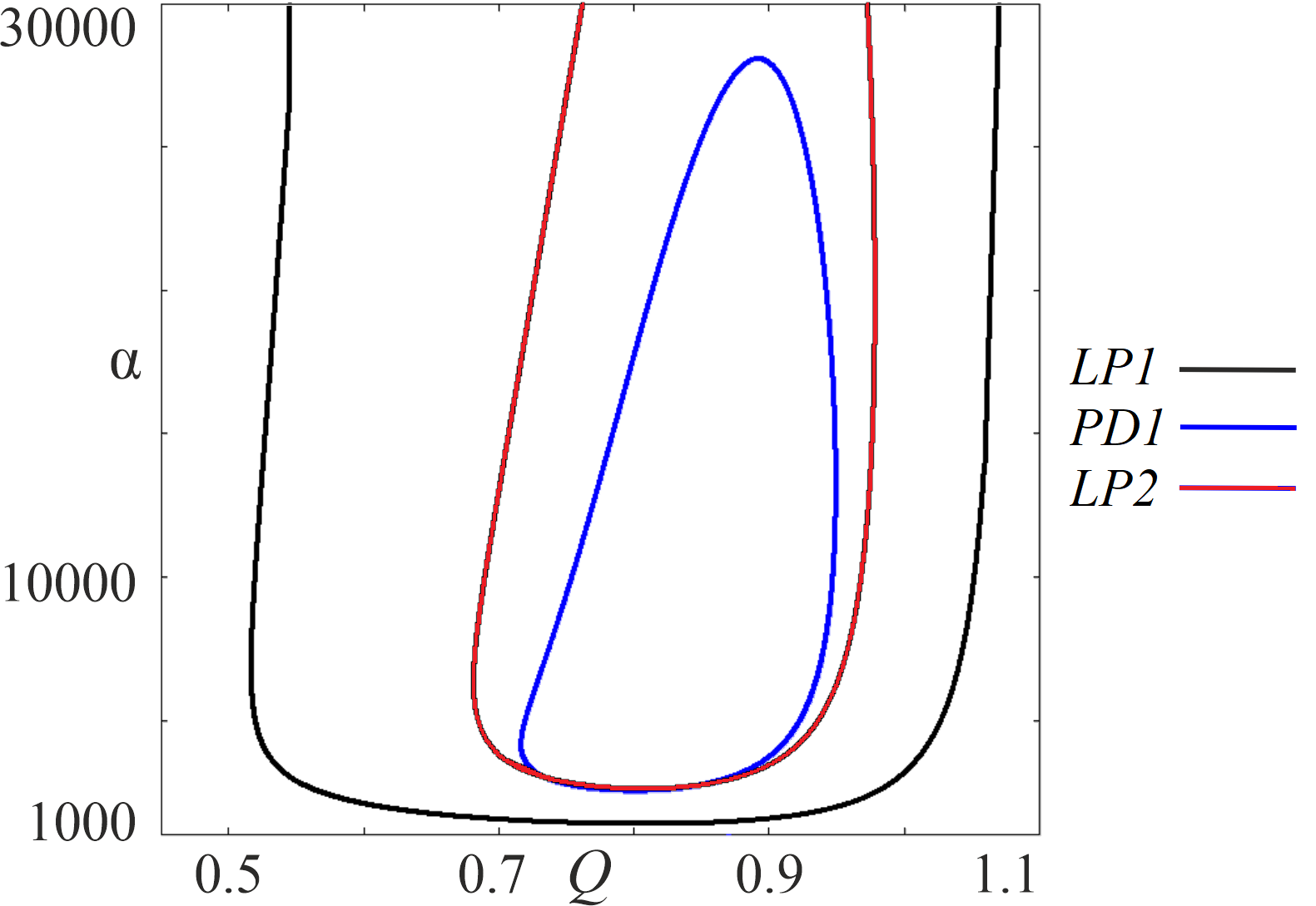}
\end{center}
\caption{Two-parameter ($\alpha$ - $Q$) continuations of LC1L2in2S bifurcations : two saddle-node LP1 (black line), LP2(red line) and period doubling PD1 (blue line)
}
\label{fig11:SR9_xpp_q_alpha}
\end{figure}

Two-parameter continuations of the main bifurcations of the chosen limit cycles are the convenient and powerful method to demarcate the parameter areas where new modes may appear. For example, two-parameter continuations of the period doubling PD1(Fig.~\ref{fig4:SR2_XPP2777}) and two saddle-node LP1 (Fig.~\ref{fig4:SR2_XPP2777}), LP2 (Fig.~\ref{fig7:SR5_XPP2777_Zoom}) bifurcation points show a wide area in the ($\alpha$ - $Q$)-plane where the regimes starting from LC1L2in2s can be expected. In Fig.~\ref{fig11:SR9_xpp_q_alpha} this diagram is presented.

There two-parameter continuations point out the presence of selected (LP and PD) bifurcations of attractors typical of $\alpha = 2777$ but they do not guarantee that the structure of one-parameter continuations, the stability of attractors and the set of attractors do not vary with $\alpha$. To clarify the dynamics, it is necessary to observe the set of regimes by direct numerical simulations and/or calculation of one-parameter continuations because these methods are working with the attractors as a whole rather than with their bifurcation points.

According Fig.~\ref{fig11:SR9_xpp_q_alpha}, a decrease in $\alpha$ causes the stable LC1L2in2s to disappear via the merging of saddle-node boundary bifurcations. Direct integration confirm this tendency demonstrating the hyperchaos dominance in accordance with Fig.~\ref{fig2:chartLEs}.
For $\alpha > 2777$ the possible parameter area with LC1L2in2s is still large, which opens the ways for other dynamics. The qualitative changes of dynamics are manifested as loss of LC1L2in2s stability in the $Q$-interval over the left part of the map in Fig.~\ref{fig2:chartLEs} for $\alpha > 4000$.  

\begin{figure*}
\includegraphics[width=0.7\textwidth]{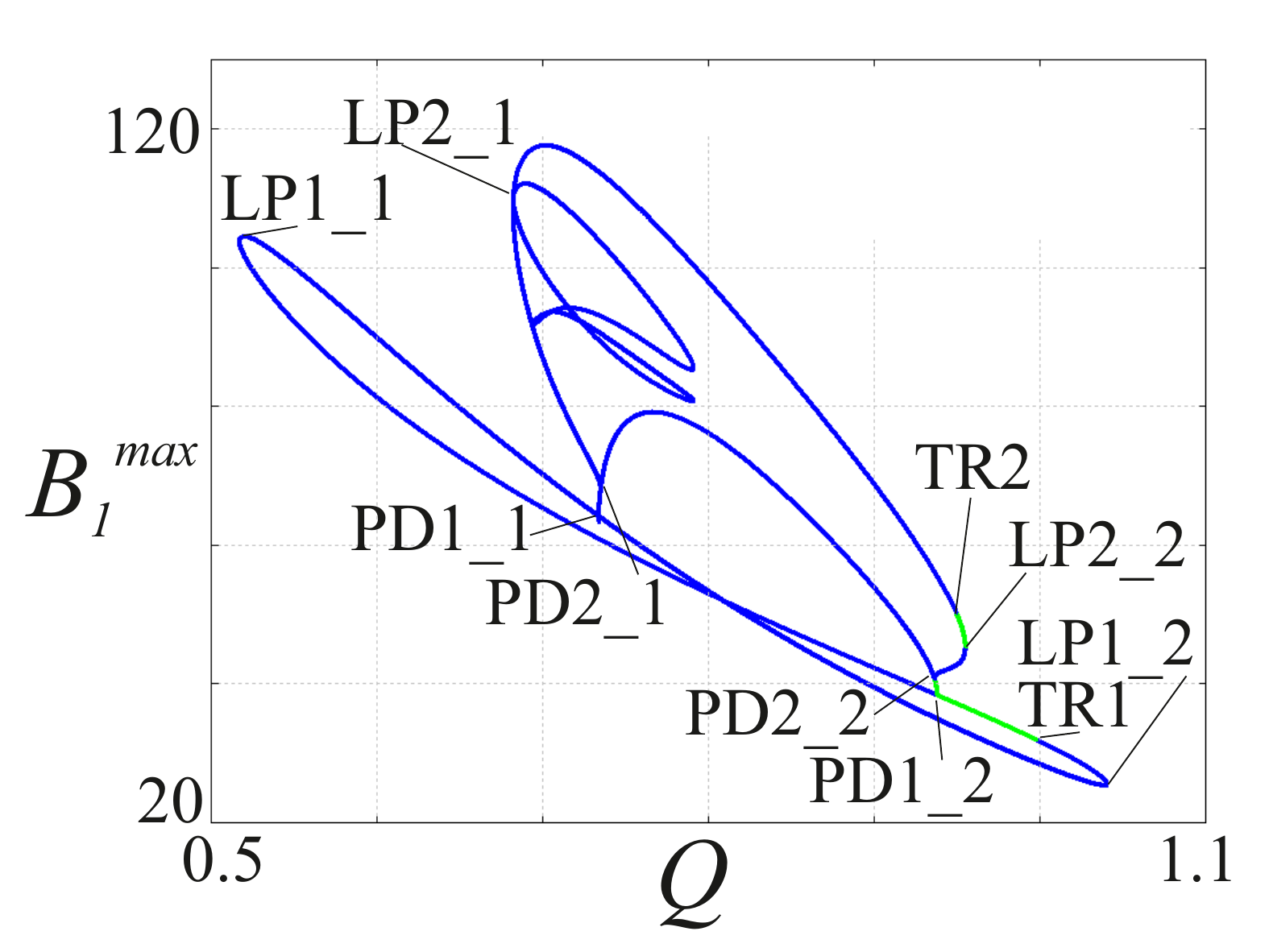}
\caption{The bifurcation continuations of inhomogeneous limit cycles starting from LC1L2in2s for $\alpha = 7000$: the lines corresponding to stable (unstable) regimes are green (blue), respectively. LP1\_1 - $Q = 0.517$; LP2\_1 - $Q = 0.682$; PD1\_1 - $Q = 0.7334$; TR1 - $Q = 1.005$; PD2\_1 - $Q = 0.7351$; PD2\_2 - $Q = 0.936$; PD1\_2 - $Q = 0.9383$, TR2 - $Q = 0.95$; LP2\_2 - $Q = 0.955$
}
\label{fig12:SR12_xpp7000}
\end{figure*}

One parameter $Q$-continuation of LC1L2in2s for $\alpha = 7000$ is presented below (Fig.~\ref{fig12:SR12_xpp7000}). An increase in $\alpha$ leads to the narrowing of the $Q$-interval, where the LC1L2s2s is stable in the right part of the $Q$-continuation and to the formation of many bifurcations of unstable inhomogeneous limit cycles in the left part of the continuation.   
However, the sequence of key bifurcations that control the appearance and stability of LC1L2in2s, LC2L4in4s and LC1L4L8out8s coincides with that for $\alpha = 2777$.

\begin{figure*}
\centering
\includegraphics[width=0.9\textwidth]{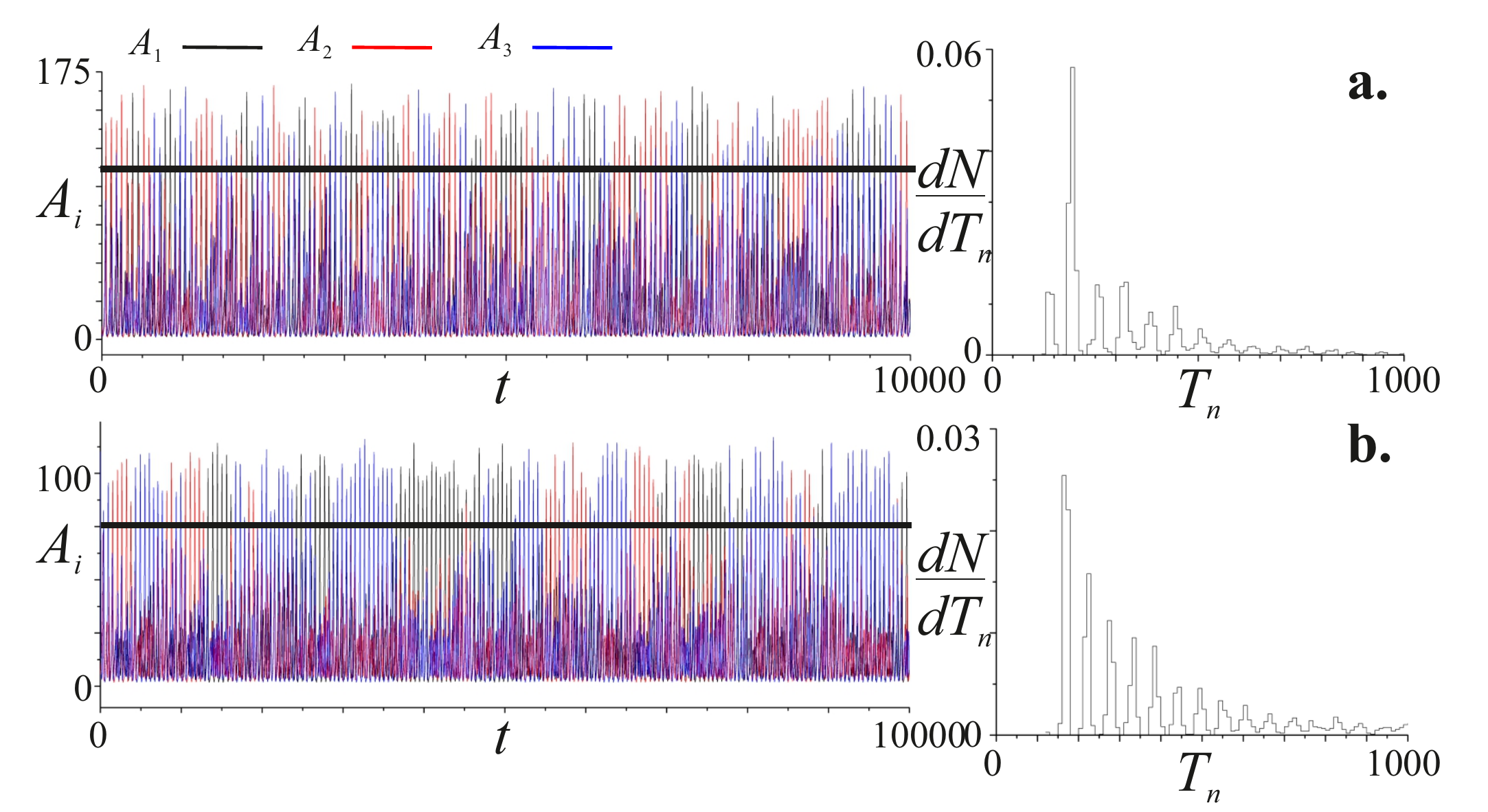}
\caption{Time series and corresponding return times ($\rm{T_n}$) distributions for $\alpha = 7000$ for $T_n > 200$.  Acr is the Poincar\'{e} section level: \textbf{a.} $Q = 0.6$, Acr=115; \textbf{b.} $Q = 0.85$, Acr=75. The grid lines show the positions of Poincar\'{e} sections
}
\label{fig13:SR11}
\end{figure*}

Although hyperchaos is the dominant attractor over almost all interval of coupling strengths, the role of the unstable inhomogeneous limit cycles is clearly seen in hyperchaotic orbits. By way of examples, we present the time series for $Q = 0.6$ and $Q = 0.85$ where there are two and three different unstable inhomogeneous limit cycles, respectively (Fig.~\ref{fig13:SR11}).  
Hyperchaos orbits include a random sequence of stripes with pieces of inhomogeneous unstable limit cycles. The closer $Q$ to the values corresponding to the inhomogeneous limit cycles stabilization (PD2\_2 at $Q = 0.936$ Fig.~\ref{fig12:SR12_xpp7000}), the smaller the deviations of the limit cycles multipliers from unity and the longer the durations of the bands filled with inhomogeneous cycles.

To calculate the distributions of these stripe widths we use Poincar\'{e} sections of the time series with fixed value Acr for variables $A_1$, $A_2$, $A_3$. The Acr values are chosen close to but somewhat lower than the variables amplitudes (as presented in Fig.~\ref{fig13:SR11}) to avoid crossing small amplitude oscillations by the Poincar\'{e} section. The comparison of the return times distributions for $Q = 0.6$ and for $Q = 0.85$ shows that the long inhomogeneous limit cycles parts in hyperchaos are significantly presented over a wide interval of the coupling strengths, being especially pronounced for strong coupling.   

\newpage

\section{Conclusions and Discussion}
\label{sec:Concl}
A lot of publications describe the appearance of collective asymmetric regimes in systems of coupled identical oscillators. The coexistence of these regimes with symmetric attractors depends on the type of oscillators and the schemes of their coupling. 
One early example was given by Tyson and Kauffman \cite{tyson1975control} who found asymmetric limit cycles in their study of two diffusely coupled identical Brusselators which, as later was found, demonstrate many other regimes \cite{volkov1995bifurcations}.
 
Since that time transitions from inhomogeneous steady state or homogeneous limit cycle to inhomogeneous limit cycle have been intensely studied in coupled identical oscillators (see \cite{banerjee2018transition, sathiyadevi2019inhomogeneous} and references therein).
Recently, we have demonstrated that there are wide regions of parameters controlling the dynamics of two and three quorum sensing coupled Repressilators where the stable inhomogeneous steady states (typically named "oscillation death") are formed, as well as inhomogeneous limit cycles born from them via Andronov-Hopf bifurcation \cite{ullner2008multistability, koseska2010cooperative}.
Here \cite{hellen2020emergence}, we explore a new type of inhomogeneous limit cycle recently found in two QS-coupled identical Repressilators after their amplitudes were increased. This cycle originates from the pitchfork bifurcation of the unstable in-phase limit cycle, has winding number 1:2 and very different oscillation amplitudes.  It dominates phase space over significant intervals of model parameters and coexists with many symmetric limit cycles and a chaotic regime.     

In the system of three QS-coupled Repressilators, hyperchaos is the dominant regime if the coupling strength is greater than some critical value \cite{stankevich2021chaos}. However, we demonstrate that a family of strongly asymmetric collective regimes exist over wide intervals of the coupling strength $Q$ and $\alpha$ for a given set of other model parameters. Hyperchaos is replaced by one of the stable attractor from this family as the model parameters are varied. In any case the final attractor is stable LC1L2in2s which turns to LC2L4in4s via period doubling bifurcation. Further evolution of the dynamics is different from the well-known Feigenbaum cascade because the next bifurcation not only doubles the oscillation period but also initiates a new inhomogeneous cycle LC4L8out8s with nonidentical orbits with small amplitudes. Both cycles are stable and coexist over short parameter intervals converting to inhomogeneous chaos or returning to hyperchaos outside these intervals. 
The existence of these inhomogeneous attractors in the parameter window surrounded by hyperchaos manifests itself in the hyperchaos "skeleton" as the formation of asymmetric parts (pieces) in the timeseries. They are often observed not only in the long transients to stable inhomogeneous cycles (see e.g. Fig.~\ref{fig5:SR3_TSs}) but even in stationary hyperchaotic orbits for the parameter intervals where all inhomogeneous cycles are unstable (see e.g. Fig.~\ref{fig13:SR11}b).
 
 As discovered recently, chimera states can exist in the system of small number identical oscillators \cite{ashwin2015weak, maistrenko2017smallest, ebrahimzadeh2020minimal}. In particular, minimal chimeras were found in all-to-all coupled modified Kuramoto phase oscillators with inertia \cite{maistrenko2017smallest} and in phase-lag coupled three metronomes \cite{ebrahimzadeh2020minimal}. These interesting regimes manifest themselves as a mismatch of average frequency between two synchronous and one desynchronized oscillator. In contrast, inhomogeneous limit cycles in QS-coupled Repressilators have the same period and the asymmetry arises due to very different amplitudes.  
It is worth mentioning  that LC1L2in2s is different from the standard partial-in-phase \cite{collins1994group, yoshimoto1993coupling, vanag2000oscillatory} limit cycle typically observed in three repulsively coupled identical oscillators. 
In this paper, we show that three simple 3-variables identical ring oscillators coupled by mean-field diffusion of signal molecules, which are produced proportionally to the concentration of one variable and activate the production of the downward variable, give interesting inhomogeneous limit cycles, quasiperiodic and chaotic oscillations with two identical orbits over significant areas of the parameter plane. 
We suggest that these results may stimulate the development of similar designs for oscillators interactions focused on generating regimes with broken symmetry.

\bibliographystyle{spphys}
\interlinepenalty=10000

\section*{Statements and Declarations}

\textbf{Funding}
Numerical simulations were implemented with financial support of the Russian Science Foundation (grant No. 21-12-00121).

\smallskip\noindent
\textbf{Competing Interests} The authors declare that they have no conflict of interest.

\smallskip\noindent
\textbf{Author Contributions} All authors whose names appear on the submission made substantial contributions to the conception of the work, the acquisition, analysis, and interpretation of data, as well as the creation of new software used in the work; drafted the work and revised it critically for important intellectual content; approved the version to be published; and agree to be accountable for all aspects of the work in ensuring that questions related to the accuracy or integrity of any part of the work are appropriately investigated and resolved.

\smallskip\noindent
\textbf{Data Availability} The data used to support the findings of this study are included within the
article.

\end{document}